\documentclass[letter]{aa}  

\usepackage{graphicx}
\usepackage{txfonts}

\usepackage[breaklinks,colorlinks=true,citecolor=black,linkcolor=black,urlcolor=blue,bookmarks=true]{hyperref}
\usepackage{lipsum}
\usepackage{subcaption}         
\usepackage{lscape}             
\usepackage{placeins}   
                                
\newcommand{\kms}{\mbox{km\,s$^{-1}$}}
\newcommand{\cz}{cz_\mathrm{LG}}

\begin{document}

\title{Edge-on galaxies relative to edge-on view of the Local Supercluster}

\author{
P.\ Dolgosheeva\inst{1}
\and 
D.\ Makarov\inst{2}
\and
N.\ Libeskind \inst{3}
}

\institute{
Saint Petersburg State University, Saint Petersburg, Russia 
\and 
Special Astrophysical Observatory, Russian Academy of Sciences, Nizhnij Arkhyz, 369167 Russia
\and
Leibniz Institut f\"{u}r Astrophysik Potsdam (AIP), An der Sternwarte 16, D-14482, Potsdam, Germany
}

   \date{Received September 30, 20XX}

 
\abstract{
Cosmological theories suggest that the angular momentum of galaxies should be closely linked to the structure of the cosmic web.
The Local Supercluster is the closest and most studied structure where the orientation of galaxy spins can be studied.
As noted by \citet{2004ApJ...613L..41N}, the use of edge-on galaxies greatly simplifies this task by reducing it to an analysis of the distribution of position angles in the Supergalactic coordinates.
We reexamine this correlation using modern catalogs that allow us to perform a more robust statistical analysis.
We test the dependence on redshift, spatial position, luminosity, and color.
We find that the spins of galaxies with stellar mass $M_*<10^{8.7}$~$M_\sun$ show a weak tendency to be aligned perpendicular to the plane of the Local Supercluster at the 2-sigma level.
Other subsamples do not show statistically significant correlations.
}


\maketitle

\section{Introduction}

According to modern cosmological concepts, the angular momentum (spin) is a fundamental property of a galaxy and contains information about the formation and evolution of stellar systems. 
The idea that tidal forces are the origin of the galaxy rotation dates back to the works of \citet{1934ApJ....79..460S,1951pca..conf..195H}.
The tidal-torque theory developed by \citet{1969ApJ...155..393P,1970Afz.....6..581D,1984ApJ...286...38W} describes the acquisition of angular momentum by a proto-galaxy in the early Universe due to tidal forces from neighboring proto-galaxies and large-scale structures.
The main prediction of the tidal-torque theory is that galaxy spins should be nearly perpendicular to the minor axis of the shear tensor associated with the distribution of the surrounding matter.
In other words, the spin of a galaxy must lie in the plane of the wall of the large-scale structure to which it belongs~\citep{2004ApJ...613L..41N}.
This occurs because in the frame of the Shear tensor's eigenvectors, the three components of angular momentum depend on the difference in the eigenvalues in an analytical way, namely $J_i =e_{ijk} (\lambda_j-\lambda_k)$, where $J_i$ is the angular momentum in the $i$ direction,  $e_{ijk}$ is the Levi-Cevita symbol and $\lambda_i$ is the eigenvalue associated with that direction. 
As such, the angular momentum in the $J_2$ direction is maximized (since $\lambda_3-\lambda_1$ is by definition greatest).

There are a number of studies that confirm the theoretical picture~\citep{1994MNRAS.271...19G, 2013MNRAS.428.1827T,2019MNRAS.485.2492C,2020MNRAS.492..153B}.
Moreover, modern mass surveys allow one to detect more subtle effects.
Using the Sloan Digital Sky Survey (SDSS), \citet{2013ApJ...775L..42T} found the difference in orientation of elliptical and spiral galaxies.
The minor axis of the ellipticals is predominantly perpendicular to the host filaments, while the spin of spirals aligns with it.
Using integral-field kinematics from the SAMI survey across the GAlaxy and Mass Assembly (GAMA) fields, \citet{2020MNRAS.491.2864W} detected a mass-dependent flip in spin orientation: the spin of low-mass galaxies tends to be along the filament, while the spin of more massive galaxies is preferentially orthogonal to the filament.
Based on 3D spins from the MaNGA survey, \citet{2021MNRAS.504.4626K} found a morphological segregation in galaxy orientation: late-type, predominantly low-mass galaxies are aligned, while S0-galaxies are perpendicular.
It was also found that the orientation is related to the stellar mass of the bulge.
Galaxies with low-mass bulge tend to parallel their spins with filament, whereas high-mass bulge galaxies are more perpendicularly aligned~\citep{2022MNRAS.516.3569B}. 

On the other hand, a number of studies do not confirm the existence of spin alignment with respect to large-scale structure or find only an extremely weak correlation~\citep[see for example][]{2016MNRAS.457..695P,2019ApJ...876...52K,2021AstBu..76..248A}. 
In particular, \citet{2023MNRAS.522.4740K} studied the angular momenta of 720 galaxies enclosed in the Local Volume of 12~Mpc around the Milky Way.
Among them, a sample of 27 elite spirals with angular momenta exceeding 0.15 that of the Milky Way stands out. 
They contribute more than 90\% of the total angular momentum of the Local Volume.
Nevertheless, their spin distribution in the sky does not show a preferred alignment with respect to the Local Sheet plane.
Therefore, the question of galaxy spin alignment relative to the large-scale structure requires further study.

The Local Supercluster appears to be an ideal laboratory to test this hypothesis.
This nearest supercluster of 30--40 Mpc diameter is one of the most studied structures in the Universe.
Due to the location of our Galaxy inside the Local Sheet, the Local Supercluster clearly stands out in the distribution of nearby galaxies in the sky as a narrow belt with a noticeable concentration of galaxies around the center in the Virgo cluster.
This fact simplifies the identification of the Local Supercluster plane, which is determined by the great circle of the Supergalactic coordinate system.
There is a wealth of literature on the study of its structure, composition, and kinematics \citep[see for example][]{2011MNRAS.412.2498M,2016ApJ...833..207K,2018AstBu..73..124K}. 

\citet{2004ApJ...613L..41N} pointed out that the use of edge-on galaxies greatly simplifies the analysis.
In the case of arbitrarily oriented disk galaxies, one needs to know the positional angle, inclination, and side of the disk that is closer to the observer.
For edge-on galaxies, the position angle (PA) is measured with high accuracy, the inclination is fixed at $90^\circ$ with an error less than $5^\circ$, and the position of the spin in the plane of the sky eliminates the ambiguity with the orientation.
Moreover, in the case of the Local Supercluster, this further simplifies the task, reducing it to measuring the PA of the major axis of the edge-on galaxy in the Supergalactic coordinates.
Using this approach, \citet{2004ApJ...613L..41N} discovered that the nearby edge-on disk galaxies are predominantly oriented perpendicular to the Local Supercluster plane, which is consistent with numerical models of the formation of dark matter halo angular momentum within the large-scale structure. 
This effect is clearly visible in the histogram of Supergalactic PAs for edge-on spiral galaxies with $cz<1200$~\kms{} \citep[see left panel of fig.~2 in the original paper by][]{2004ApJ...613L..41N}.
However, this conclusion was based on a sample of only 30 edge-on galaxies from the Principal Galaxy Catalog \citep[PGC,][]{1997A&AS..124..109P}.

Thanks to massive deep photometric and redshift surveys over two decades since the work by~\citet{2004ApJ...613L..41N}, the sample of known Local Supercluster galaxies has increased by at least an order of magnitude.
In addition, new catalogs of edge-on galaxies~\citep{2014ApJ...787...24B, 2022MNRAS.511.3063M} with well-defined selection criteria have recently been published.
All these allow us to use the approach proposed by~\citet{2004ApJ...613L..41N} with much better statistics and to test the alignment of galaxy spins with respect to the plane of the Local Supercluster at a new level.

\section{Sample}

\begin{table}
\centering
\caption{The sample sizes}
\label{tab:StatsOnSamples}
{\small
\begin{tabular}{lcc@{\;}c@{\;}c@{\;}c}
\hline\hline
Sample  & 
\multicolumn{2}{c}{$\cz<3600$} & $<$1200 & 1200--2400 & 2400--3600 \\
\cline{2-3}
&
&
phot\\
\hline
EGIPS & 552 & 508 &        73 & 225 & 254 \\
EGIS  & 88 &  82 &        28 &  38 &  22 \\
LEDA  &1049 & 934 &       129 & 341 & 579 \\
\hline
Total &1689 &1524 &       230 & 604 &855 \\
\hline\hline
\end{tabular}
}
\end{table}

Our work is based on the catalog of edge-on galaxies~\citep[EGIPS,][]{2022MNRAS.511.3063M} found in the Pan-STARRS1 survey~\citep{2016arXiv161205560C}, which is the largest list of edge-on galaxies to date, containing 16551 objects.
The EGIPS catalog covers three quarters of the sky above $\delta=-30^\circ$.
Due to the structure of the public Pan-STARRS1 survey and the neural network used to search for galaxies, the EGIPS catalog may undercount the largest galaxies, as well as galaxies of low surface brightness.
In order to populate these gaps, we used the edge-on disk galaxies from the SDSS catalog~\citep[EGIS,][]{2014ApJ...787...24B} containing 5749 objects and a sample of galaxies from the HyperLeda database~\citep{2014A&A...570A..13M}. 

The EGIPS and EGIS catalogs specialized in the study of edge-on galaxies, but galaxies from the HyperLeda required additional preparation efforts.
HyperLeda galaxies were selected according to the following parameters: \texttt{vlg<3600 and t>-3 and mod0>25 and inc>85},
where \texttt{vlg} is the object velocity, $\cz$, in the reference frame of the Local Group centroid~\citep{1996AJ....111..794K}; \texttt{t} is the de Vaucouleur's numerical code of the morphological type of the galaxy; \texttt{mod0} is a distance modulus to exclude the Local Group members; \texttt{inc} is the inclination estimated from the observed galaxy axial ratio and a characteristic thickness depending on its morphology.
Measurement errors, intrinsic variation in characteristic thickness, and misclassifications in morphology can lead to large errors in the inclination estimates.
To avoid this problem, we visually inspected all selected galaxies to rule out the wrong cases.
In the final sample, we included only spiral galaxies because the inclination angle of S0 galaxies is difficult to determine.

The statistics on the subsamples of edge-on galaxies inside different redshift ranges with respect to the centroid of the Local Group~\citep{1996AJ....111..794K} are gathered in Table~\ref{tab:StatsOnSamples}.
The rows count only unique objects added to the previous catalogs.
The last row gives the total number of galaxies in each subsample.
The third column, phot, indicates the number of galaxies with good photometry.
To analyze the alignment, \citet{2004ApJ...613L..41N} used only 30 galaxies with velocities less than 1200~\kms{}.
As can be seen in Table~\ref{tab:StatsOnSamples}, the new data sample has increased almost 8 times.

\section{Analysis}

\begin{figure*}
\centering
\begin{tabular}{c@{}c}
\includegraphics[width=0.5\linewidth]{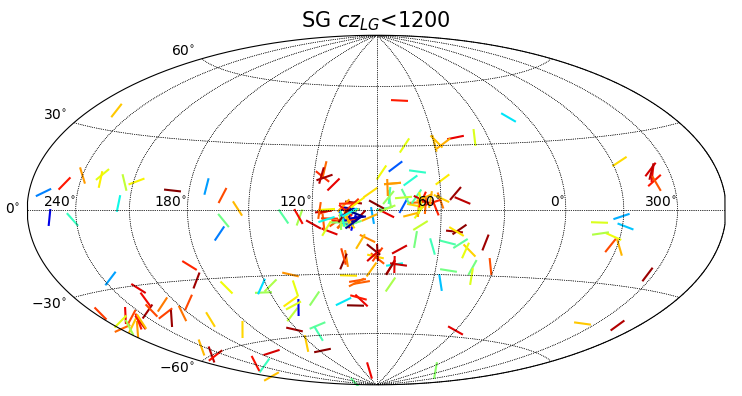} &
  \includegraphics[width=0.5\linewidth]{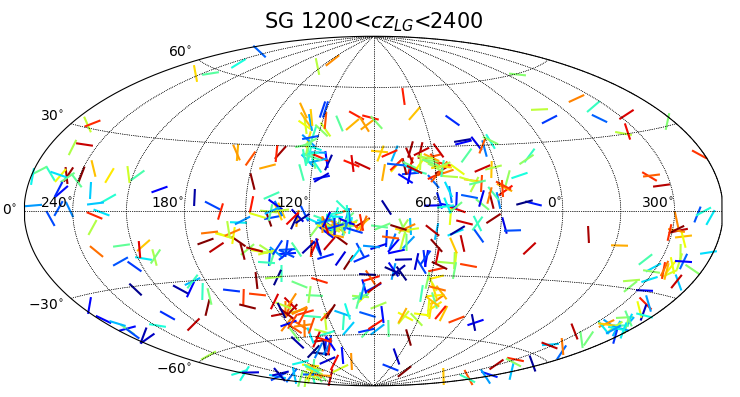} \\
\includegraphics[width=0.5\linewidth]{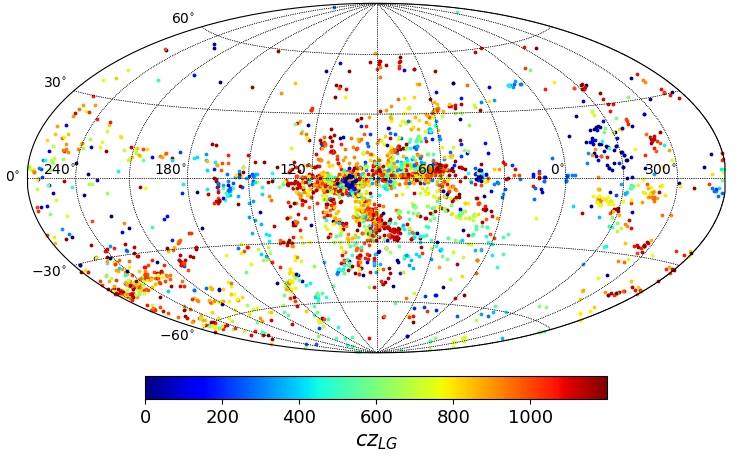} &
  \includegraphics[width=0.5\linewidth]{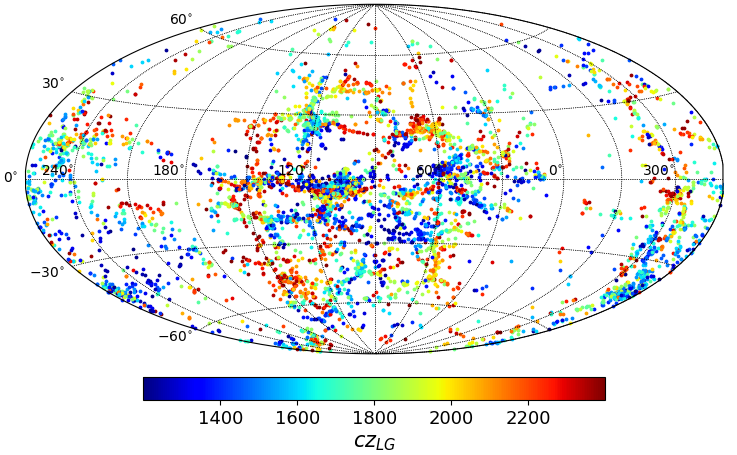} \\
\end{tabular}
\caption{\label{fig:edgeonSkyMap1200}
All-sky maps of galaxy distribution in Supergalactic coordinates (Aitoff projection).
\textbf{Top row:} Edge-on galaxies in our sample, split into two redshift intervals: $\cz<1200$~\kms{} (left) and $1200<\cz<2400$~\kms{} (right).
Major axes are shown as short segments.
\textbf{Bottom row:} Nearby galaxies from the 50 Mpc Galaxy Catalog~\citep{2024AJ....167...31O} in the same redshift intervals as above.
Colors indicate redshifts in the Local Group centroid reference frame.
}
\end{figure*}

To analyze the PA distribution of the edge-on galaxies in the Supergalactic coordinates, we use the standard equations of spherical trigonometry to convert PA from the J2000.0 equatorial system to the Supergalactic one.
In our analysis, the PA is adjusted to the range $[0^\circ, 90^\circ)$, since the orientation at an angle $\alpha > 90^\circ$ relative to the Supergalactic plane is equivalent to an angle of ($180^\circ - \alpha) < 90^\circ$.

The all-sky distribution of the edge-on galaxies with indication of their orientations is plotted in the Supergalactic coordinates in the top panels of Fig.~\ref{fig:edgeonSkyMap1200}. 
For clarity, we show two layers in separate panels:
nearby galaxies with $\cz\leq1200$~\kms{} (top left panel), where we expect to see the alignment effect most clearly, 
and more distant ones in the velocity range $1200<\cz\leq2400$~\kms{} (top right panel). 
For convenience, the bottom panels of Fig.~\ref{fig:edgeonSkyMap1200} map the distribution of arbitrarily oriented galaxies in the same velocity range, based on the modern 50~Mpc Galaxy Catalog~\citep{2024AJ....167...31O}.
The panels show a complex network of the large-scale structure in the nearby Universe at different scales.
The centers of the figures correspond to the Supergalactic coordinates $(l_\mathrm{SG},b_\mathrm{SG}) = (90^\circ, 0^\circ)$, which is quite close to the center of the Local Supercluster in the Virgo cluster $\sim(102.9^\circ, -2.3^\circ)$.

The nearest redshift region $\cz\leq1200$~\kms{} is dominated by the Local Supercluster and most galaxies are concentrated in a narrow belt along the Supergalactic Plane.
In addition, the Leo Spur, a structure at $b_\mathrm{SG}\approx-15$ parallel to the Local Supercluster Sheet, and the near periphery of the Fornax Supercluster at $\sim (262.5^\circ,-42.1^\circ)$ are well distinguished. 

To our surprise, the edge-on galaxies in Fig.~\ref{fig:edgeonSkyMap1200} do not show expected prominent perpendicular orientation in the nearest $\cz<1200$~\kms{} layer either with respect to the Local Supercluster or Leo Spur plane.

\begin{figure}
\centering
\includegraphics[width=\linewidth]{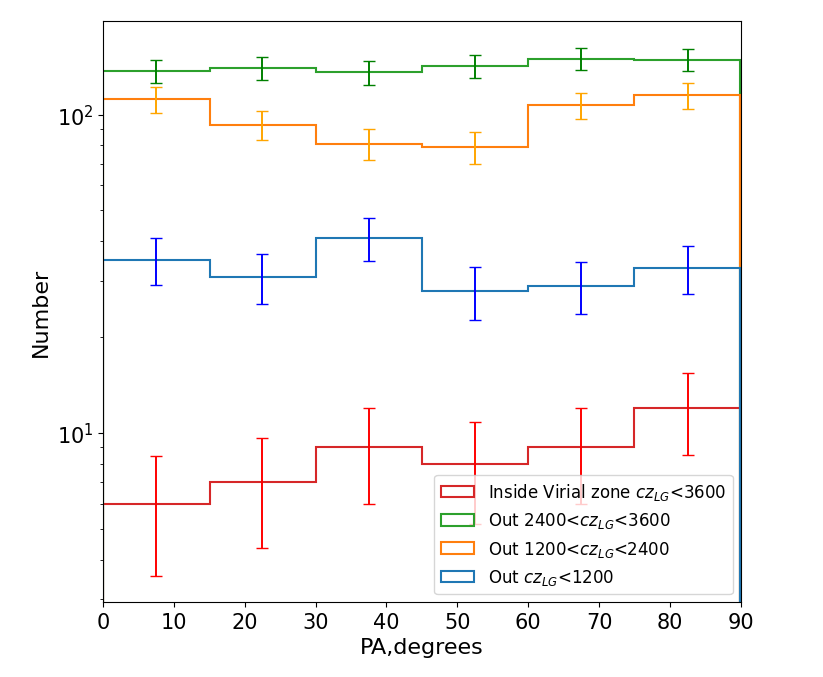}
\caption{
Supergalactic PA distribution of edge-on galaxies from our final sample inside and outside the virial zone of the Virgo Cluster.
} 
\label{fig:PAczVirgo}
\end{figure}

For the quantitative analysis, we considered 4 subsamples.
Because galaxies in the virial zone of a cluster should be randomized in velocities, positions, and orientations, in order to purify the statistics of their influence, we excluded from consideration the Virgo cluster galaxies lying within the $6^\circ$ radius of M~87 with velocities $\cz<3600$~\kms{}.
We split the remaining galaxies into 3 layers by redshift: $\cz\leq1200$, $1200<\cz\leq2400$ and $2400<\cz\leq3600$~\kms{}.
The PA distribution of these subsamples is shown in Fig.~\ref{fig:PAczVirgo}.
The main impression is that the distribution of galaxies across PA is fairly uniform and does not show any statistically significant dominance in orientations.
It appears that the galaxy disks within the virial zone tend to lie in the plane of the Local Supercluster, but the Kolmogorov-Smirnov test shows that this alignment is statistically insignificant with a $p$-value of 0.466.

Obviously, samples of galaxies with a simple redshift selection criterion are clogged with neighboring structures.
To improve the selection of galaxies belonging to the Local Supercluster, we consider only objects lying within the $\pm2$ and $\pm5$~Mpc layers from the Local Supercluster plane.
As before, the virial zone of the Vigro cluster is excluded from consideration.
Using photometry data from the 50~Mpc Galaxy Catalog~\citep{2024AJ....167...31O}, we can separate galaxies by morphology and luminosity.
The distribution of edge-on galaxies by $(g-i)_0$ color shows a pronounced bimodality, and the color $(g-i)_0=0.86$ splits them into two roughly equal subsamples of blue and red galaxies.
We also divide the sample by the stellar mass $M_*=10^{8.7}$~$M_\sun$ into two approximately equal parts of bright and faint galaxies.

\begin{table}
\centering
\caption{
Asymptotic $p$-value of the Kolmogorov-Smirnov test of the null hypothesis that the PA distribution is uniform for different subsamples of edge-on galaxies. 
The $p$-value $<0.07$ are marked in bold.
The sample size is given in parentheses.
}
\label{tab:kstest}
{\small
\begin{tabular}{lcccc}
\hline\hline
Sample & \multicolumn{2}{c}{$\cz\leq1200$~\kms{}}   & \multicolumn{2}{c}{$\cz\leq2400$~\kms{}} \\
       & $\pm2$~Mpc      & $\pm5$~Mpc            & $\pm2$~Mpc            & $\pm5$~Mpc          \\
\hline
all    &  0.468 (68)     & 0.852 (115)           & 0.405 (124)           & 0.223 (239)\\
\hline
bright &  0.204 (32)     & 0.260 \phantom{0}(46) & 0.448 \phantom{0}(60) & 0.339 (111)\\
faint  &  \textbf{0.062} (31)     & 0.279 \phantom{0}(61) & \textbf{0.024} \phantom{0}(48) & \textbf{0.041} (109)\\
red    &  0.264 (25)     & 0.216 \phantom{0}(41) & 0.448 \phantom{0}(48) & 0.370 \phantom{0}(91) \\
blue   &  0.422 (38)     & 0.586 \phantom{0}(66) & 0.199 \phantom{0}(60) & 0.161 (129)\\
\hline\hline
\end{tabular}
}
\end{table}

\begin{figure*}
\centering
\includegraphics[width=0.49\linewidth]{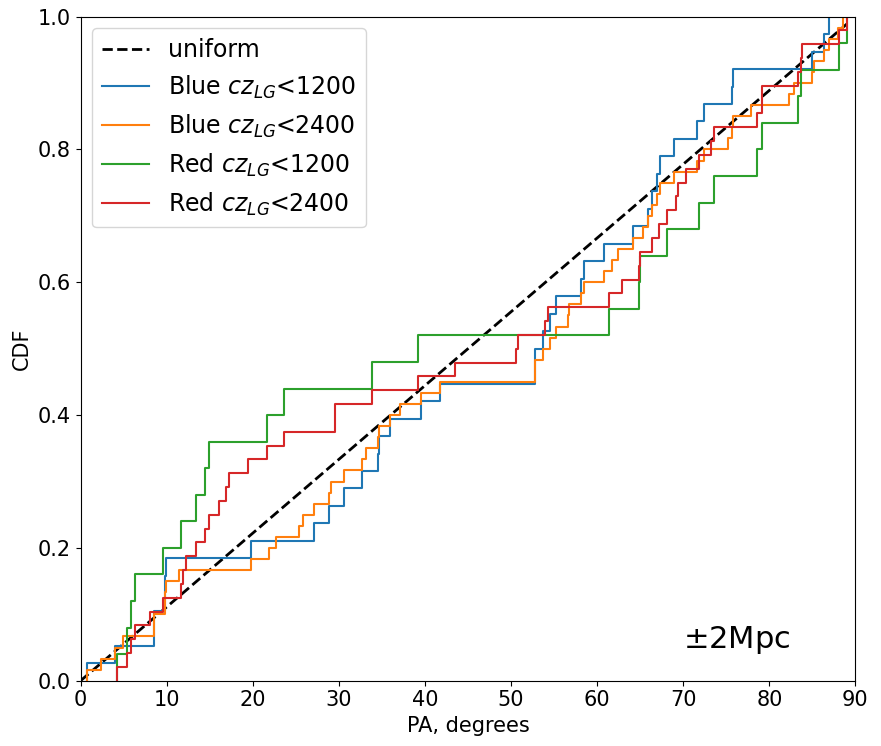}
\includegraphics[width=0.49\linewidth]{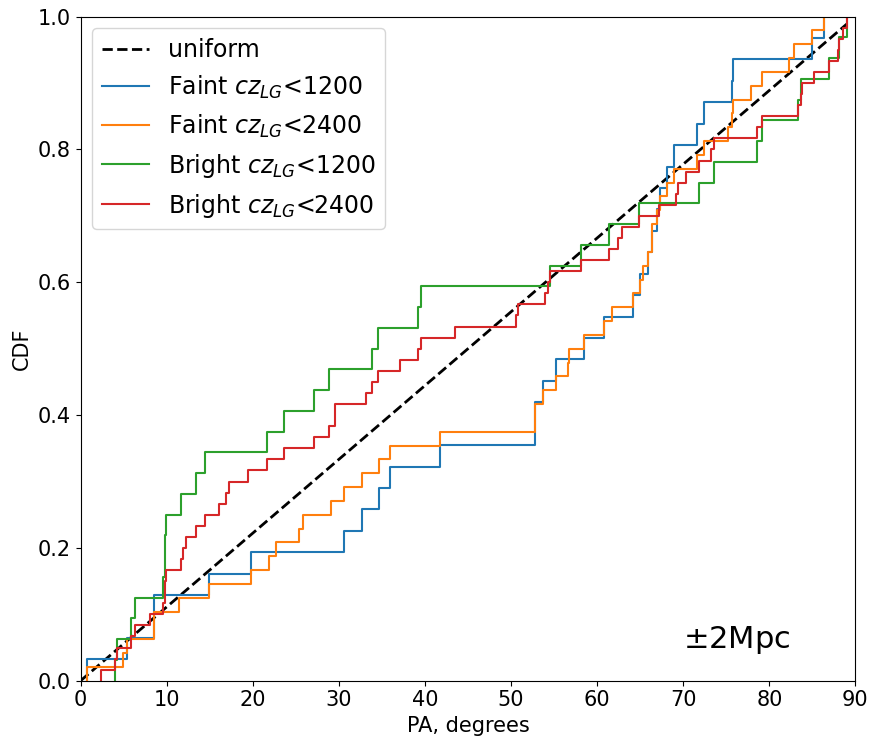}
\includegraphics[width=0.49\linewidth]{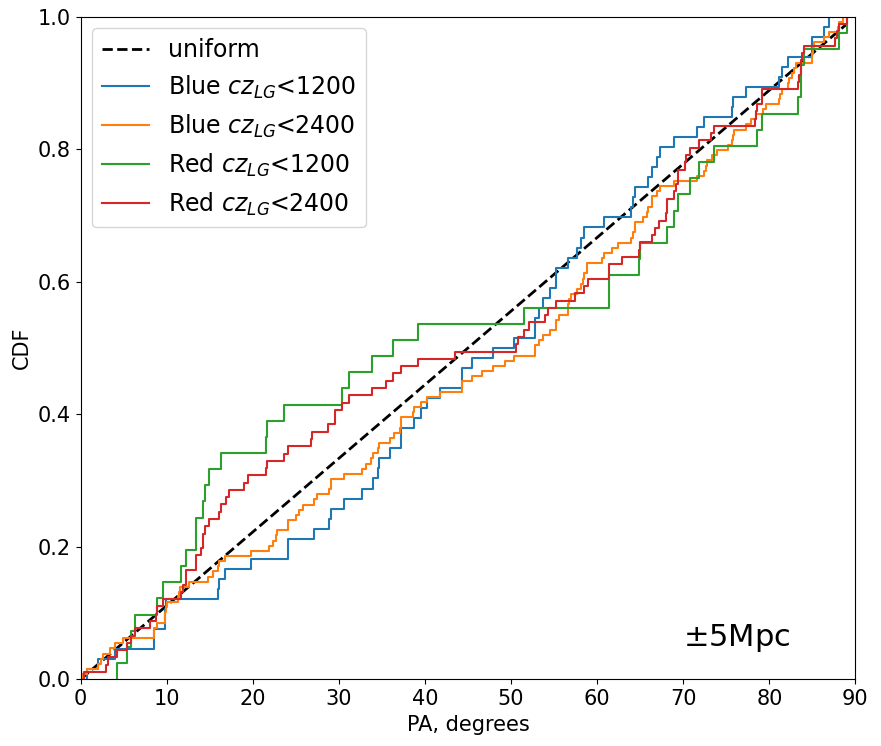}
\includegraphics[width=0.49\linewidth]{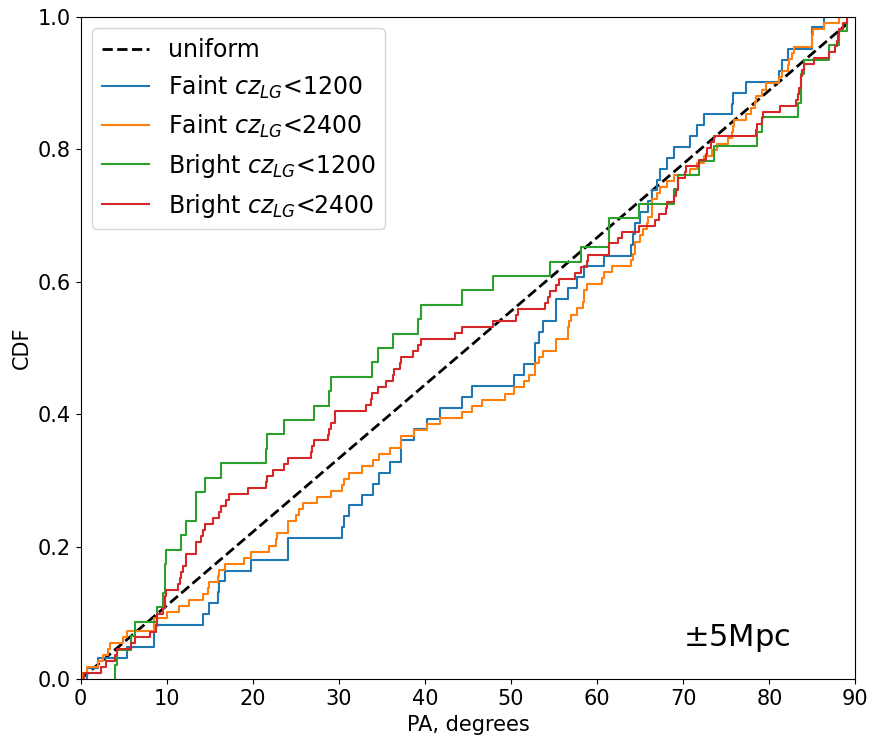}
\caption{
Cumulative PA distributions of edge-on galaxies with respect to the Supergalactic plane.
The panels show the distributions within the  $\cz<1200$ and $\cz<2400$~\kms{} ranges for subsamples of the bright and the faint galaxies, separated by stellar mass at $M_*=10^{8.7}$~$M_\sun$, as well as for the blue and the red galaxies, divided by color at $(g-i)_0=0.86$~mag. The \textbf{top row} corresponds to the $\pm2$~Mpc layer relative to the Supergalactic plane, while the \textbf{bottom row} shows the distributions within the $\pm5$~Mpc layer.
The virial zone of the Virgo cluster is excluded from consideration.
The black dashed line indicates the uniform distribution.
}
\label{fig:cdf}
\end{figure*}

We perform the one-sample Kolmogorov-Smirnov test under the null hypothesis that PAs are uniformly distributed.
The cumulative distributions of Supergalactic PA for subsamples of different morphologies and luminosities, measured in two regions of different redshift depths, are shown in Fig.~\ref{fig:cdf}.
The results are summarized in Table~\ref{tab:kstest}.
The tests showed that the Supergalactic PA distributions of edge-on galaxies in most of the subsamples do not differ from random at the $\alpha = 0.05$ significance level.
The only exception is the case for faint objects.
Galaxies of small stellar masses, $M_*<10^{8.7}$~$M_\sun$, show predominantly perpendicular spin orientation relative to the Local Supercluster plane with a difference from the random distribution at the significance level better than $\alpha=0.07$ in almost all cases considered.
Despite the statistical significance of the result at approximately the 2-sigma level, one could consider it a fairly robust detection, as it is observed in three out of four cases---if not for a few caveats.
These samples are not fully independent as the larger regions encompass the smaller ones.
The deviation from the null hypothesis for the subsample of nearby faint galaxies in the thick slice, $\cz\leq1200$~\kms{} \& $\pm5$~Mpc, is statistically insignificant, with its $p$-value differing sharply from the other three values, though it is comparable to those of other non-correlating subsamples.

\section{Summary}

We analyze the distribution of the galaxy orientations relative to the Local Supercluster plane, following the idea proposed by~\citet{2004ApJ...613L..41N} using a sample of 1689 nearby edge-on disk galaxies within $\cz < 3600$~\kms{} compiled from the EGIPS~\citep{2022MNRAS.511.3063M} and EGIS~\citep{2014ApJ...787...24B} catalogs, and the HyperLeda database~\citep{2014A&A...570A..13M}.
It is about an order of magnitude larger than in the original article.

We test different subsamples selected by redshift, Local Supercluster sheet membership, color, and stellar mass.
Unlike \citet{2004ApJ...613L..41N}, we do not find any statistically significant alignment of spins to the Local Supercluster plane either among the nearest galaxies, $\cz<1200$~\kms{}, or in any other subsample.
The distribution of the Supergalactic PAs of edge-on galaxies appears to be random, except for a subsample of faint galaxies.
The disks of low stellar mass, $M_*<10^{8.7}$~$M_\sun$, belonging to the Local Supercluster show a tendency to be aligned in its plane.
In other words, their spins are oriented predominantly perpendicular to the plane of the Local Supercluster.
This effect is observed within the $\pm2$ and $\pm5$~Mpc layers relative to the Supergalactic plane, in the regions of $\leq1200$ and $\leq2400$~\kms{} in redshift space, at a significance level of 2.4--6.2\%, which corresponds to 2.3--1.9 sigma.
This result contradicts the expectation that the spin of low-mass galaxies is aligned along the filaments found in the works of \citet{2020MNRAS.491.2864W} and \citet{2021MNRAS.504.4626K}.
It should be noted that in the subsample of faint galaxies within $\pm5$~Mpc and $\cz<1200$~\kms{} this correlation breaks down.
Given the relatively small sample size, it cannot be ruled out that this is a consequence of some statistical fluctuation.

The lack of spin alignment to the plane of the Local Supercluster predicted by theory may lie in the fact that galaxies are part of different filaments and are subject to complex tidal influences from the surrounding large-scale structure.
Thus, although tidal force fields should lead to spin alignment, the intersections of filaments (knots) and the presence of sheets create complex time-varying gravitational fields that can lead to random orientation of spins.
The interactions at these junctions can disrupt any initial alignment. 

The explanation for the lack of any correlations of spins and the Local Supercluster plane can lie in the upcoming paper of \citet{2024arXiv241218880A}. 
The authors quantitatively estimate the possible spatial and time evolution of the potential of the Milky Way analogs in the HESTIA cosmological simulations~\citep{2020MNRAS.498.2968L} and the role of satellite galaxies and the Local Group in the formation of the potential of the Galaxy. 
One of the conclusions is that the axis of rotation of the galactic disk can slowly change its orientation by an angle of up to 70 degrees. The authors do not know the reason for this behavior. 

\begin{acknowledgements}
This work was supported by the Russian Science Foundation grant \textnumero~24--12--00277. 

We acknowledge the usage of the HyperLeda database\footnote{\url{http://leda.univ-lyon1.fr}}~\citep{2014A&A...570A..13M}.
\end{acknowledgements}

\bibliographystyle{aa}
\bibliography{ref}


\end{document}